# Intertwined Swirling Polarization States in BaTiO$_3$ with Embedded BaZrO$_3$ Nanoregions

R. Machado,[1] F. Di Rino,[1,2] M. Sepliarsky,[1] and M. G. Stachiotti[1†]

[1]*Instituto de Física Rosario, UNR-CONICET, 27 de Febrero 210 Bis, 2000 Rosario, Argentina.*
[2]*Institute of Physics of the Czech Academy of Sciences, Na Slovance 2, 18200, Prague 8, Czech Republic.*

Ferroelectric materials embedded with dielectric inclusions offer a unique platform for exploring novel topological polar textures. Using first-principles-based atomistic simulations, we investigate the polarization behavior of a BaTiO$_3$ matrix containing segregated BaZrO$_3$ nanoregions. We demonstrate that the polar texture in three-dimensionally ordered arrays of dielectric inclusions is governed by their size and spacing, revealing three distinct regimes. At large separations, the nanocomposite exhibits bulk-like BaTiO$_3$ phase transitions, while at smaller spacings, interconnected swirling polarization patterns give rise to vortex supercrystal states. We analyze the stabilization mechanisms of these states and show that each regime is characterized by distinct switching behavior. Furthermore, we find that nanocomposites with randomly distributed dielectric inclusions exhibit swirling polarization textures, giving rise to an amorphous network of entangled vortices. Our findings provide new insights into the physics of relaxor ferroelectrics, are consistent with recent experimental observations, and open up new possibilities for designing materials with emergent topological functionalities.

The field of ferroelectricity has recently undergone a paradigm shift. While traditional research focused on bulk perovskites, recent breakthroughs have demonstrated ferroelectricity in single-element 2D monolayers [1,2] and, conversely, the emergence of complex topological textures—such as vortices, skyrmions, and hopfions—in oxide nanostructures [3–10]. These discoveries have attracted significant interest due to their potential in nanoelectronics, data storage, and sensing, where precise control over polarization states could fundamentally enhance device functionality.

In recent years, substantial progress has been made in understanding the interplay between dielectric and ferroelectric materials at the nanoscale, particularly in paraelectric-ferroelectric superlattices and nanocomposites comprising ferroelectric nanostructures embedded within a dielectric matrix [11–16]. Interfacial interactions in these systems give rise to unique domain structures, resulting from the competition among electrostatic, elastic, and polarization-mediated forces. From a theoretical perspective, computational methods like the effective Hamiltonian approach have been used to study these systems [17,18]. Furthermore, phase-field modeling has predicted the formation of interconnected Hopfion-like excitations in disordered ensembles of contacting ferroelectric nanoparticles [19].

Despite extensive efforts on ferroelectric-dielectric nanocomposites, the inverse system—dielectric nanostructures embedded in a ferroelectric matrix—is largely unexplored. This system holds significant technological potential, particularly given its relevance to relaxor ferroelectrics like BaZr$_x$Ti$_{1-x}$O$_3$ (BZT), where Zr segregation into Zr-rich regions influences relaxor behavior [20]. We investigate the polar behavior of these nanocomposites, specifically dielectric BaZrO$_3$ (BZ) nanoregions embedded in a ferroelectric BaTiO$_3$ (BT) matrix, using first-principles-based atomistic simulations. Our findings show that small dielectric inclusions trigger significant polarization rearrangements, leading to the emergence of novel intertwined swirling polarization textures. These topological patterns are controllable by adjusting the size and spacing of the inclusions.

The predicted polar textures were obtained from molecular dynamics (MD) simulations [21] on supercells of varying sizes under periodic boundary conditions, using interatomic potentials previously developed for BaTiO$_3$ [22] and BaZrO$_3$ [23] from a first-principles-based atomistic approach [24,25]. Details of the computational method are described in the End Matter section. The interatomic potentials for BT and BZ have proven highly accurate in reproducing the thermal polar behavior of both pure materials [22,23] and BZT solid solutions [26], exhibiting good agreement with experimental data.

We first consider a three-dimensional array of BZ nanoregions in the form of cubic inclusions embedded in a BT ferroelectric matrix. A specific array is defined by the pair ($d$, $s$), where $d$ represents the lateral dimension of the BZ nanocubes and $s$ denotes their separation, both expressed in lattice constant units.

We studied the thermal evolution of polar configurations for inclusions with lateral dimension d=5 and varying separation s. We used two order parameters: total polarization (**P**) and the toroidal moment per inclusion (**G**). The results identified three distinct regimes. In the first regime, characterized by large separations (low Zr concentration), the nanocomposite shows the same phase transition sequence as bulk BaTiO$_3$. For example, the (5,12) array (approximately 2.5% Zr concentration) exhibits the characteristic R–O–T–C sequence of BT, with almost identical transition temperatures (Fig.1a).

Figures 1b–d depict the polarization patterns of the room-temperature tetragonal phase in the (5,12) array. The BT matrix shows almost uniform polarization along the tetragonal axis, but this is disrupted near the inclusions, where local polarization vectors reorient, resembling laminar fluid flow around an obstacle. Although the BZ nanoregions have zero net polarization, interfacial atomic relaxations [27,28] induce small local polarization vectors directed toward the inclusion center (Fig. 1b). This polar configuration is an antihedgehog-like topological defect (a hedgehog ball with converging polarization [29]), similar to structures observed in BZT simulations [30] and Bi$_{0.5}$Na$_{0.5}$TiO$_3$-based relaxors [31].

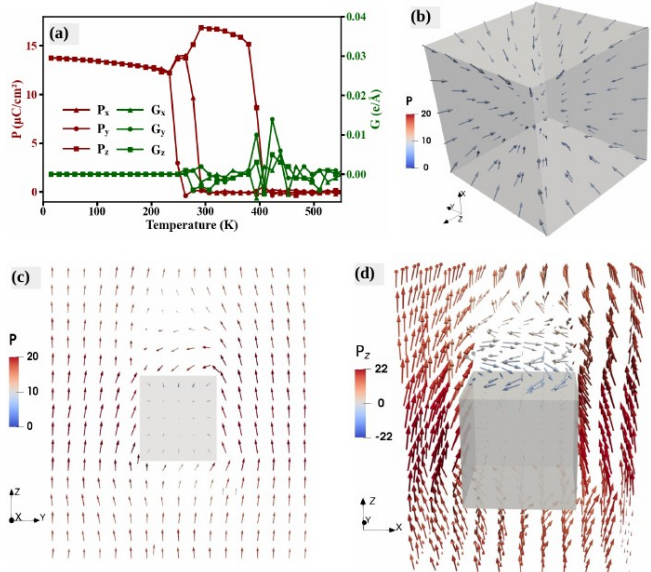

FIG 1. **(a)** Temperature dependence of polarization **P** and toroidal moment **G** for the (5,12) array, showing the bulk BaTiO$_3$ phase sequence. **(b-d)** Room-temperature polarization pattern, showing **(b)** hedgehog-like structure within the BaZrO$_3$ nanoregion, **(c)** a uniform polarized BT matrix, and **(e)** the subtle whirlpool effect (vortex) near the interface.

A small vortex-like structure emerges near the interface perpendicular to the macroscopic polarization (Fig. 1d). This structure, characterized by swirling local polarization vectors, creates a subtle whirlpool effect. The overall polar behavior of the nanocomposite thus resembles laminar flow around an array of 'sink-like' obstacles, an analogy with fluid dynamics [10] that remains valid even at higher concentrations of dielectric inclusions.

Next, we investigate the impact of reducing the spacing between dielectric inclusions. Our results show that smaller separations favor the stabilization of interconnected swirling polarization patterns at the dielectric nanoregions, giving rise to vortex supercrystal (VSC) states. Figure 2a displays the temperature dependence of **P** and **G** for the (5,9) array, corresponding to a Zr concentration of 4.5%.

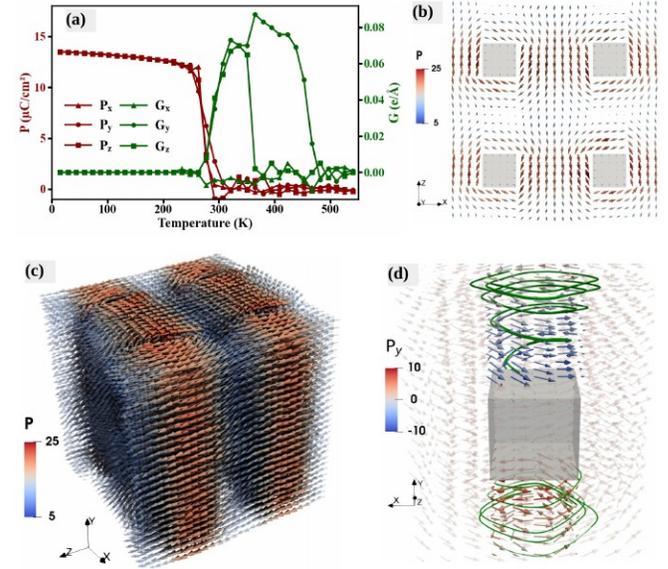

FIG 2. **(a)** Temperature dependence of **P** and **G** for the (5,9) array, stabilizing vortex-like phases. **(b-c)** Polarization pattern of the room-temperature V$_2$ phase, illustrating the two-dimensional VSC state characterized by y-axis tubular vortices. **(d)** Close-up near an inclusion, showing two interconnected swirling polarization structures with opposite chiralities, driven by the 'sink-like' nature of the inclusion.

At low temperatures, the nanocomposite exhibits a rhombohedral phase similar to that of the (5,12) array. However, this phase does not evolve into homogeneously polarized orthorhombic or tetragonal phases. Instead, both phases fragment into domains through the stabilization of vortex-like states with zero net macroscopic polarization. As shown in Fig. 2a, two vortex-like phases are stabilized: one with nonzero $G_y$ and $G_z$ components, and the other, at higher temperature, with only the $G_y$ component remaining nonzero.

The polarization patterns of the latter phase (Figs. 2b–c) reveal a VSC state, characterized by tubular vortices rotating around the y-axes passing through the inclusions. This phase features alternating clockwise and counterclockwise vortices along the x-axis, preserving their rotational sense along the z-direction. The BT regions between inclusions polarize along the tetragonal z-axis with alternating ±z orientations, resulting in 180˚ tetragonal ferroelectric domains (Fig. S1

[32]). This configuration is the $V_2$ phase, where each inclusion hosts two independent spiraling vortices that converge at the dielectric nanoregion. As shown in Fig. 2d, the $V_2$ phase exhibits spiral-like polarization structures merging near the interface. The inclusion's 'sink-like' character drives a whirlpool effect, creating two interconnected swirling structures with opposite chiralities (left- and right-handed vortices). The low-temperature state, referred to as $V_4$, is an extension of $V_2$ where vortices rotate around two axes (y and z), characterized by four independent spiraling vortices converging at each nanoregion, leading to orthorhombic ferroelectric domains (Fig. S2 [32]).

Further reducing the inclusion spacing stabilizes a three-dimensional VSC state, termed the $V_6$ phase, where each dielectric inclusion hosts six independent swirling vortices. For the (5,5) array (Zr concentration of 12.5%), this phase is stabilized from 0K to ∼650K, characterized by zero net macroscopic polarization and non-zero values for all three components of the toroidal moment (Fig. 3a).

The room-temperature pattern (Fig. 3b-d) features intertwined polar vortices with toroidal moments along all three Cartesian axes, forming a 3D mosaic structure. Fig. 3b illustrates the vortex structure rotating around the y-axis that crosses one inclusion, with similar configurations observed for rotations around the x- and z-axes. Closely spaced, 'sink-like' obstacles induce a whirlpool effect in all spatial directions (Fig. 3d). A comparable three-dimensional pattern has been reported in phase-field simulations of periodic nanopores in strontium titanate under strain [33].

Analysis of the interstitial region at the center of eight inclusions—a cubic space of 5×5×5 lattice units—reveals the formation of a small vortex structure rotating around the (11-1) axis (Fig. 3e). This texture resembles the polarization pattern found in isolated BT nanocubes [34], providing insight into the complex interaction between vortices rotating around different inclusions.

We now investigated the impact of varying the lateral dimensions (d) of the inclusions using MD simulations for d=2,3,4, and 6 at different separations (s). The results are summarized in the phase diagram shown in Fig. 4. The diagram, expressed as a function of Zr concentration, is shown in Fig. S3 [32]. For d=2, the nanocomposites exhibit the bulk BT phase sequence regardless of s. However, when d>2, all three regimes are observed: the R–O–T–C sequence at large separations, and the vortex-like phases ($V_2,V_4,V_6$) stabilized as inclusion concentration increases. Crucially, the range of separations (s) over which the vortex phases are stabilized increases with the inclusion size (d). For instance, for d=3, $V_2$ and $V_4$ are stable only at s=4, and $V_6$ at s<3. For d=6, $V_2$ and $V_4$ are stable for s between 10 and 12, and $V_6$ is stable for s<10.

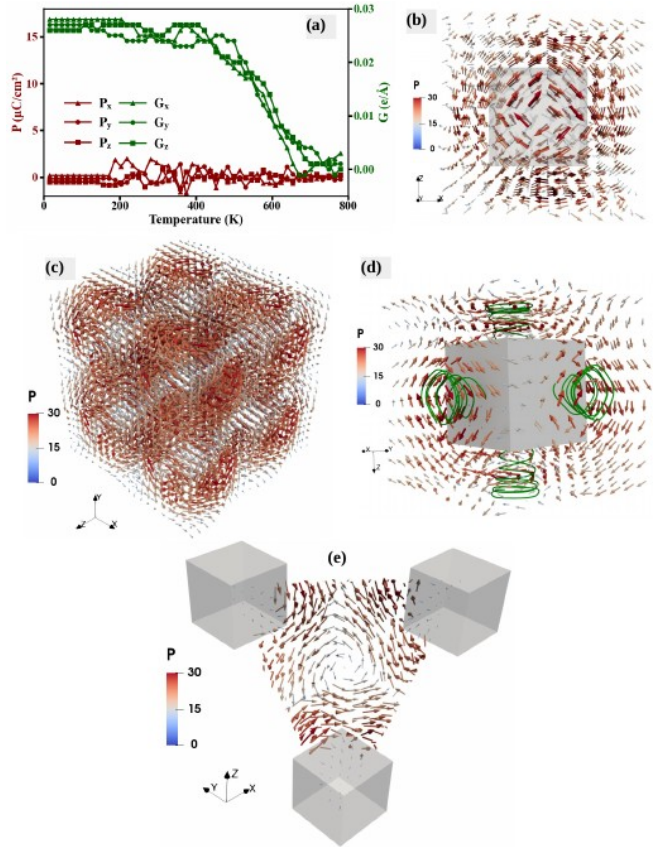

FIG 3. **(a)** Temperature dependence of **P** and **G** for the (5,5) array. Polarization pattern of **(b)** the y-axis rotating vortex structure and **(c)** the three-dimensional VSC ($V_6$) state. **(d)** Illustration of the six independent swirling vortices converging at each inclusion. **(e)** Polarization profile in the interstitial region at the center of eight inclusions, revealing a small vortex structure rotating around the (11−1) axis.

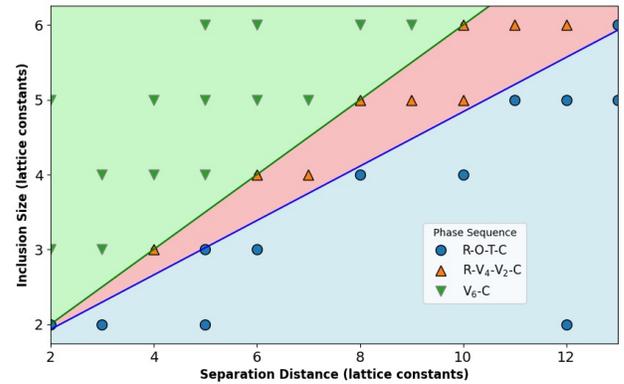

FIG 4. Phase diagram showing the polar behavior of the $BaTiO_3$ matrix with embedded $BaZrO_3$ nanoregions as a function of inclusion lateral size (d) and separation distance (s). The three distinct phase transition regimes discussed in the text are indicated by different colors. The symbols correspond to the MD simulations performed.

Additionally, parallel studies were conducted using spherical BZ nanoregions. Remarkably, the observed polar behavior is identical to that described for cubic nanoregions.

As an example, the polarization pattern of the $V_6$ phase stabilized in the array of BZ nanospheres with a diameter of d = 7, separated by a distance of s = 7, is shown in Fig. S4 [32].

The three regimes presented in the phase diagram of Fig. 4 can be distinguished by their switching behavior. To investigate this, MD simulations under an external electric field applied along the x-axis were performed to compare the polarization reversal mechanisms. The calculated P–E hysteresis loops for three representative cases with d=5 are shown in Fig. 5. The loop corresponding to the room-temperature phase of the (5,12) array (Fig. 5a) exhibits a rectangular shape similar to that of bulk $BaTiO_3$ [22], indicating sharp polarization switching. In contrast, the loop for the $V_2$ phase of the (5,9) array is more rounded (Fig. 5b), suggesting that the presence of the domain walls that traverse the inclusions modifies the switching mechanism. In fact, the butterfly-shaped hysteresis loop observed in the $G_y$ component of the toroidal moment, shown in Fig. 5b, indicates that the vortex $V_2$ phase emerges as an intermediate state in the switching process when the system adopts a zero-polarization configuration at the coercive field.

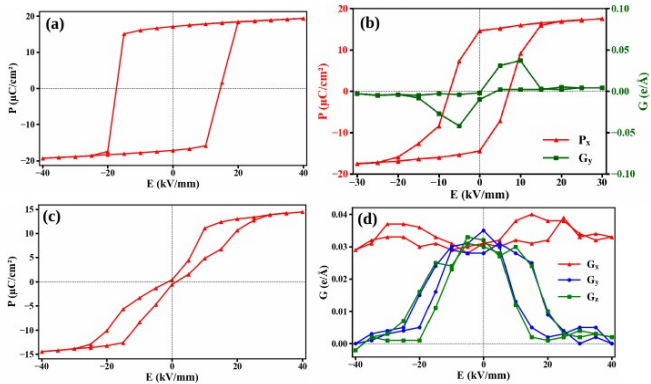

FIG 5. Molecular dynamics simulations of P-E hysteresis loops for the (5,12) [a], (5,9) [b], and (5,5) [c] arrays. The behavior of toroidal moment components under the applied electric field for the (5,9) [b] and (5,5) [d] arrays is also shown. The shape of the ferroelectric loop serves as a distinct fingerprint for each of the three identified regimes.

In contrast, the P–E loop for the (5,5) array differs markedly from the previous cases, exhibiting a pinched shape (Fig. 5c), typically associated with domain-wall pinning by defects. In this case, however, the pinching originates from the pinning of the vortex state by the dielectric inclusions. The behavior of the toroidal moment components under the applied field is shown in Fig. 5d. While the two components perpendicular to the field ($G_y$ and $G_z$) display butterfly-shaped loops, the component parallel to the field ($G_x$) remains nearly constant throughout the cycle. As a result, the system becomes polarized along the x-axis in the saturation regime, with the local polarization rotating around this axis. Upon removal of the field, the $V_6$ phase is fully recovered. These results demonstrate that the shape of the ferroelectric hysteresis loop provides a distinct fingerprint for each of the three regimes.

We also investigated the emergence of swirling polarization structures in nanocomposites with randomly distributed inclusions (d=5 nanocubes). Three systems were simulated with Zr concentrations of approximately 5%, 15%, and 25% (corresponding to 11, 32, and 54 inclusions in a 30×30×30 supercell). Figures 6a–c show the polarization patterns, displaying only local polarizations exceeding 20μC/cm² to highlight polar vortex formation. At 5% Zr (Fig. 6a), an intertwined vortex array resembling the ordered $V_2$ state is observed. As the concentration increases to 15% and 25% (Figs. 6b and 6c), the swirling structures increase and connect to form a disordered network of vortices. This configuration is analogous to the crystalline $V_6$ state but manifests as an amorphous network due to the random inclusion distribution. These results confirm the robustness of polar vortex structures even under strong positional disorder in ferroelectric nanocomposites.

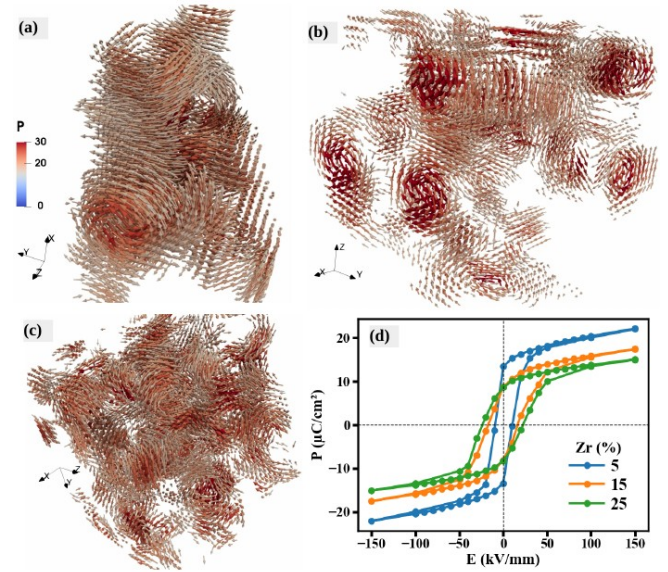

FIG 6. Polarization patterns showing swirling polarization structures in nanocomposites consisting of BZ nanoregions with a lateral dimension of d=5, randomly distributed in a BT matrix, at a Zr concentration of 5% (a), 15% (b) and 25% (c). (d) Polarization-electric field (P-E) hysteresis loop of the nanocomposites.

Figure 6d compares the hysteresis loops for the three random Zr concentrations. The 5% Zr loop closely resembles the (5,9) ordered array (Fig. 5b). The 15% and 25% Zr loops show reduced remnant polarization and increased coercive fields. Instead of the pinched loops characteristic of the ordered $V_6$ phase, these loops resemble the broader, open-loop shape seen in Fig. 5b. Importantly, the polarization profile of the poled state (non-zero macroscopic polarization at zero field) reveals swirling polarization patterns pinned by the dielectric inclusions, which is the characteristic fingerprint of the $V_6$ state.

The rich variety of swirling polar textures originates from atomic relaxation effects at the ferroelectric–dielectric interface. Since the unit cell volume of BZ is larger than BT, adjacent $TiO_6$ octahedra are deformed, inducing atomic displacements and changes in polarization in the interface region. Although BZ inclusions are initially nonpolar, these relaxations create small local polarizations directed toward the dielectric material. Thus, the dielectric inclusions act as nucleation centers for the development of swirling polarization states.

The emergence of intertwined vortex states within nanocomposites with randomly distributed dielectric inclusions is particularly relevant for relaxor ferroelectrics. X-ray absorption fine structure experiments show that Zr atoms tend to segregate into Zr-rich regions in $BaZr_xTi_{1-x}O_3$ (BZT) solid solutions [20], corresponding to $BaZrO_3$ regions that increase in size with Zr concentration. Our phase diagram (Fig. S3 [32]) suggests these Zr aggregates are sufficient to induce swirling polarization states. The remarkable similarity between the trends in our simulated hysteresis loops for random inclusions (Fig. 6d) and experimental BZT measurements [35]—which both show decreasing remnant polarization and increasing coercive field with increasing Zr content—underscores that compositional disorder shapes the ferroelectric switching behavior. Our findings, consistent with swirling textures observed in PMN-PT [36], demonstrate that compositional heterogeneities alone can induce these patterns, offering new perspectives on the role of nanoscale structures in relaxor physics.

Furthermore, our results align with recent experimental observations of conical polarization patterns in $BaTiO_3$ nanoislands grown on a dielectric $SrTiO_3$ substrate [37], where PFM data showed a spiral-like polarization structure directed toward the interface. More significantly, our findings validate the observed formation of cubic superstructures in potassium-lithium-tantalate-niobate (KLNT) systems [38–42]. These superstructures, believed to be a 3D periodic domain mosaic where each site hosts six polarization vortex cores [41,42], closely resemble our predicted $V_6$ phase. This work confirms, through first-principles-based atomistic simulations, that interactions between polar and non-polar phases drive the formation of intertwined polarization vortices, offering key insights for the rational design of nanometer-scale vortex supercrystalline phases by tailoring the size and spacing of dielectric inclusions.

In conclusion, we investigated the polar behavior of a $BaTiO_3$ ferroelectric matrix containing embedded $BaZrO_3$ nanoregions using first-principles-based atomistic simulations. Our results demonstrate that the $BaZrO_3$ inclusions act as nucleation centers for the emergence of exotic, intertwined polarization vortices. We identified three distinct regimes in three-dimensional arrays of inclusions: (i) bulk-like $BaTiO_3$ behavior when the separation between inclusions is sufficiently large, and (ii) the stabilization of vortex supercrystal (VSC) states at smaller separations, characterized by either two (and four) or six swirling polarization patterns converging at the dielectric inclusions. Each regime exhibits a characteristic hysteresis loop that serves as a fingerprint of the underlying vortex state. Additionally, we showed that randomly distributed dielectric nanoregions give rise to an amorphous network of entangled vortices, underscoring their potential relevance to the physics of $BaZr_xTi_{1-x}O_3$ relaxors. These results, together with available experimental data, indicate the presence of intertwined swirling polarization patterns in BZT, offering new perspectives on how nanoscale compositional heterogeneities influence ferroelectric behavior and the response to external fields. Our findings also offer valuable insight into recent experimental observations of vortex supercrystals in potassium-lithium-tantalate-niobate systems. Overall, this work establishes a theoretical framework for the design of dielectric–ferroelectric nanocomposites with tailored topological states, opening new avenues for applications in nanoelectronics, data storage, and photonic devices.

We are pleased to thank Igor A. Lukyanchuk and Yuri A. Tikhonov for illuminating discussions. This work was supported by Consejo Nacional de Investigaciones Científicas y Técnicas de la República Argentina (CONICET) by PIP-0374. MGS thanks support from CIUNR. FDR acknowledges the financial support of the European Union by the ERC-STG project 2D-sandwich (Grant No 101040057). Computational resources were provided by the e-INFRA CZ project (ID:90254), supported by the Ministry of Education, Youth and Sports of the Czech Republic. We also acknowledge the financial support of the 3D-TOPO HORIZON-MSCA-SE project.

**Data Availability Statement**
The data that support the findings of this study are openly available in Zenodo at https://doi.org/10.5281/zenodo.18469864. These data include the numerical datasets underlying the figures and representative output files of the simulations. Additional raw simulation outputs are available from the corresponding author upon reasonable request.

†Corresponding author stachiotti@ifir-conicet.gov.ar

# End Matter

*Appendix A: First-principles-based atomistic approach*

The approach used in this work models each ion as a charged core connected to a massless charged shell (shell model), where the equilibrium core-shell distance represents the ion's electronic polarization. Interactions between atomic species are described by interatomic potentials that include harmonic and fourth-order core-shell couplings ($k_2$ and $k_4$), long-range Coulombic forces, and short-range repulsive interactions. The model parameters for an specific material are determined by fitting first-principles calculations. This atomistic approach, combined with molecular dynamics (MD) simulations, has proven to be a powerful tool for predicting the behavior of ferroelectric compounds at finite temperatures [24,25].

The short-range interactions used in this work are modeled using two types of potentials: a Born-Mayer potential, $V(r)=Ae^{-r/\rho}$, for Ba-O, Ti-O, and Zr-O pairs, and a Buckingham potential, $V(r)=Ae^{-r/\rho}+r/C^6$, for O-O interactions. The model potential for $BaTiO_3$ [22] and $BaZrO_3$ [23] differs only in the B-cation-related interactions. Since the Zr and Ti ions are homovalent, and the O-O and Ba-O interactions are identical in both compounds, this consistency allows for the simulation of not only the pure compounds but also their solid solutions and interfaces.

Composite systems consisting of BZ nanoregions in the form of cubes and spheres embedded in a BT ferroelectric matrix were considered. Two different configurations of inclusions were investigated: a three-dimensional ordered array and randomly distributed inclusions. The first case is illustrated in Fig. 7.

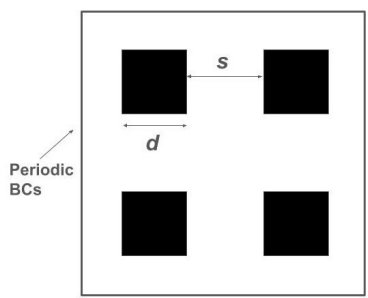

FIG 7. Schematic 2D illustration of the 3D periodic array defined by the inclusion lateral dimension d and separation s.

A specific array is defined by the pair (d,s), with each parameter expressed in units of the lattice constant. Additionally, we compared the simulation results for a given Zr concentration using two different supercells under PBC: one containing 1 dielectric inclusion and the other containing 8.

The MD simulations were performed using the DL-POLY code [21] within a constant stress and temperature (N, σ, T) ensemble. Supercells of varying sizes were employed with periodic boundary conditions (PBC). The simulations were conducted at temperature intervals of 10 K, with a time step of 0.4 fs. Each MD run consisted of at least 90,000 time steps (36 ps) for data collection, following a thermalization period of 10,000 steps (4 ps). To reduce statistical noise, the temperature dependence of the toroidal moment G was refined using an energy-window averaging technique over five adjacent temperature points (raw data are provided in Fig. S6 of the SM). Other order parameters were evaluated using standard production runs. Hysteresis loops were calculated by performing four switching cycles at each field increment ($2\times10^5$ steps), discarding the initial cycle and averaging the remaining three to ensure stable equilibrium behavior.

To ensure the ergodicity and statistical stability of the equilibrium states obtained, we selected the most representative configurations exhibiting vortex textures (specifically, the (5,9) and (5,5) arrays) and ran independent MD simulations at temperatures down to 250 K. These simulations were initiated from various distinct atomic coordinates: the perfect cubic structure, homogeneous displacements of Ti and Zr atoms along the ⟨001⟩, ⟨011⟩, and ⟨111⟩ directions, and random initial displacements. The evolution of the polarization **P** and the toroidal moment **G** (the key order parameter characterizing the vortex textures) was carefully monitored. We confirmed that, despite differences in the equilibration time, the averaged equilibrium values are statistically indistinguishable across all initial conditions. This convergence to a unique and stable equilibrium state justifies the statistical reliability of the reported averaged parameters.

Furthermore, we compared the results from heating and cooling cycles to identify the stability of low-temperature phases and the energetic competition between the uniformly polarized state and the vortex texture.

# Supplemental Material

# Intertwined Swirling Polarization States in BaTiO$_3$ with Embedded BaZrO$_3$ Nanoregions

R. Machado[1], F. Di Rino[2], M. Sepliarsky[1] and M.G. Stachiotti[1]

*1. Instituto de Física Rosario, UNR-CONICET, 27 de Febrero 210 Bis, 2000 Rosario, Argentina.*
*2. Institute of Physics of the Czech Academy of Sciences, 182 00, Praha 8, Czech Republic*

## 1. Tetragonal ferroelectric domains in the V₂ state

(a) 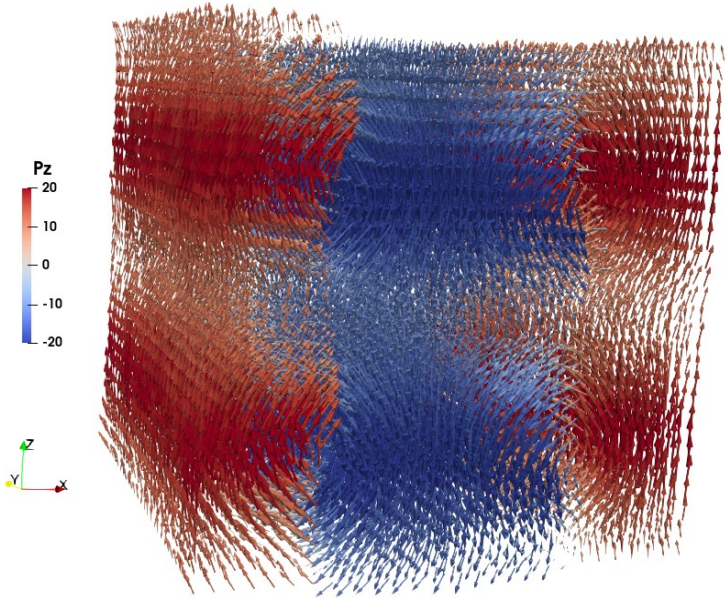

(b) 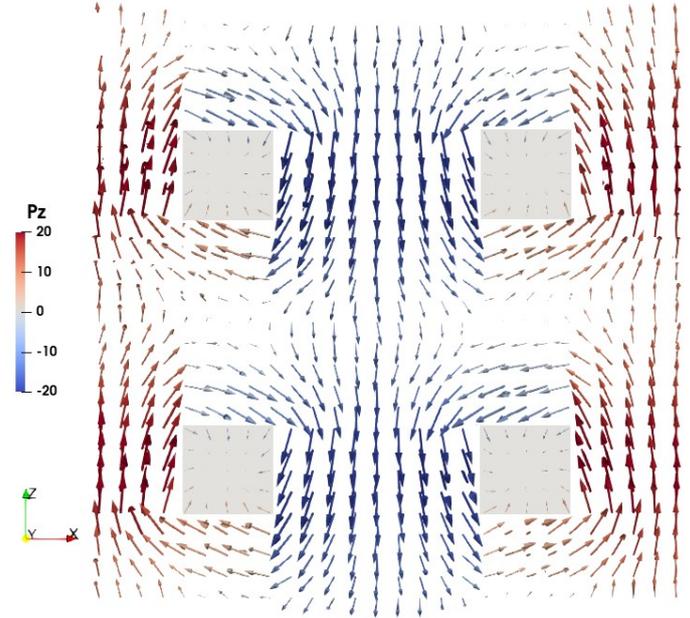

**FIG. S1:** Polarization pattern of the V$_2$ state stabilized in the (5,9) array, illustrating the formation of tetragonal ferroelectric domains. (a) Polarization configuration for a supercell containing eight BZ inclusions. (b) Two-dimensional cross-section in a plane perpendicular to the y-axis. Laminar ferroelectric domains, separated by 180° domain walls that intersect the inclusions, are clearly observed in both panels through the color map of the polarization component $P_z$. The polarization magnitude within each domain is not uniform, but rather modulated by the presence of the inclusions.

## 2. Orthorrombic ferroelectric domains in the V₄ state

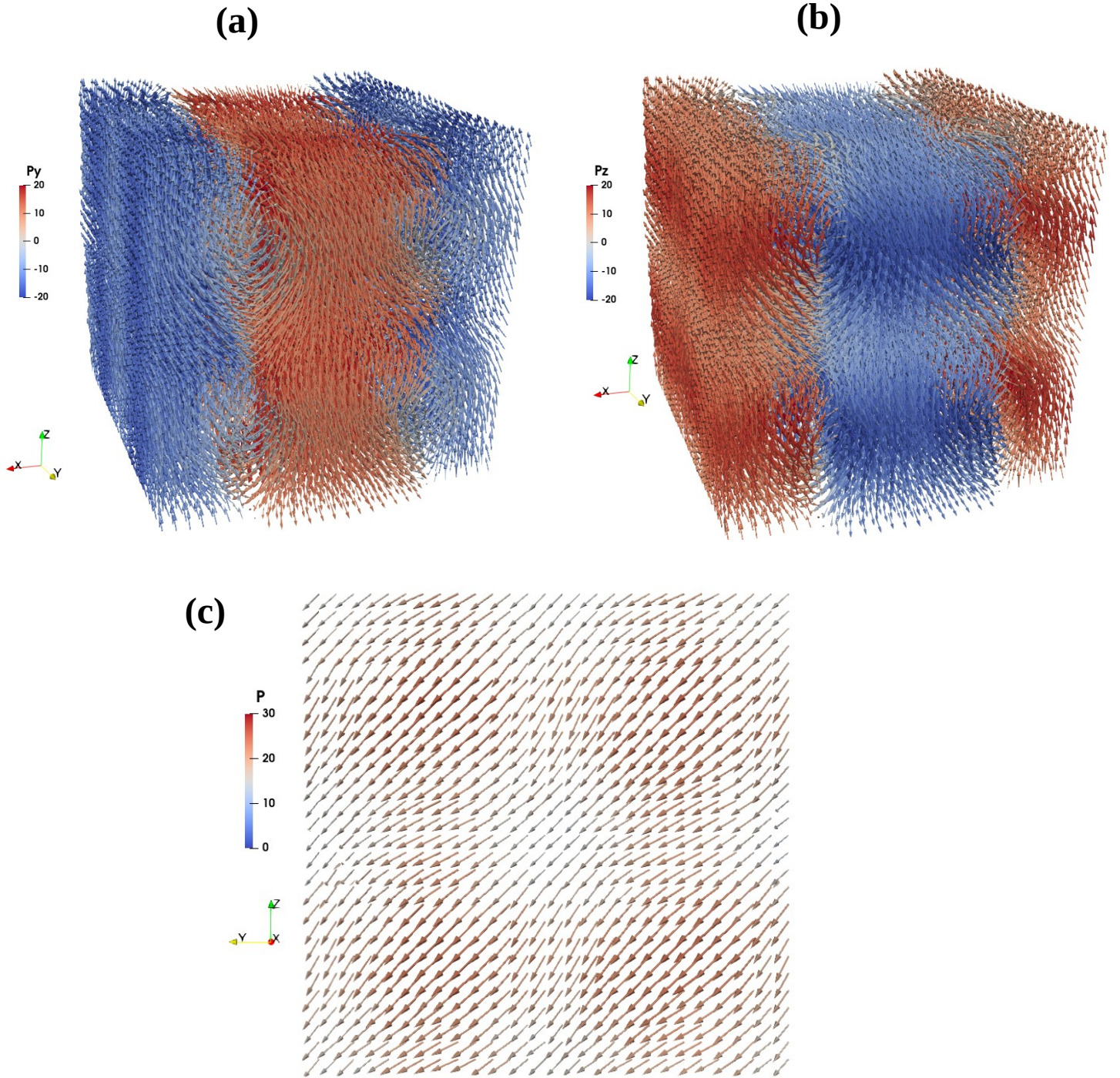

**FIG. S2:** Polarization pattern of the V₄ state stabilized in the (5,9) array, illustrating the formation of orthorhombic ferroelectric domains polarized along the (011) directions. These laminar domains, separated by domain walls that intersect the inclusions, are clearly observed in the colormaps of the $P_y$ (a) and $P_z$ (b) polarization components. (c) Two-dimensional cross-section in a plane perpendicular to the x-axis, taken through the center of a laminar domain polarized along the (0 1 $\bar{1}$) direction.

## 3. Phase diagram as a function of Zr concentration

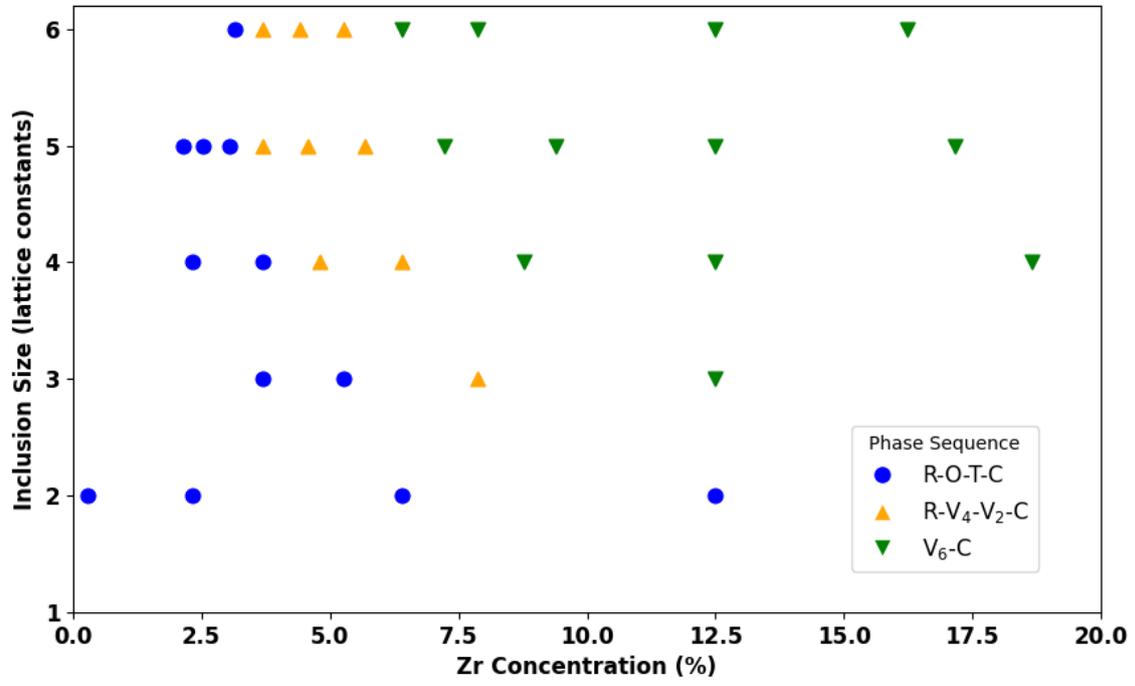

**FIG. S3**: Polar behavior of a BaTiO$_3$ matrix with embedded BaZrO$_3$ nanoregions as a function of the lateral size of the dielectric inclusions and the Zr concentration. The three distinct regimes discussed in the text are indicated by different symbols.

## 4. V$_6$ state in a BaTiO$_3$ matrix embedded with spherical BaZrO$_3$ nanoregions

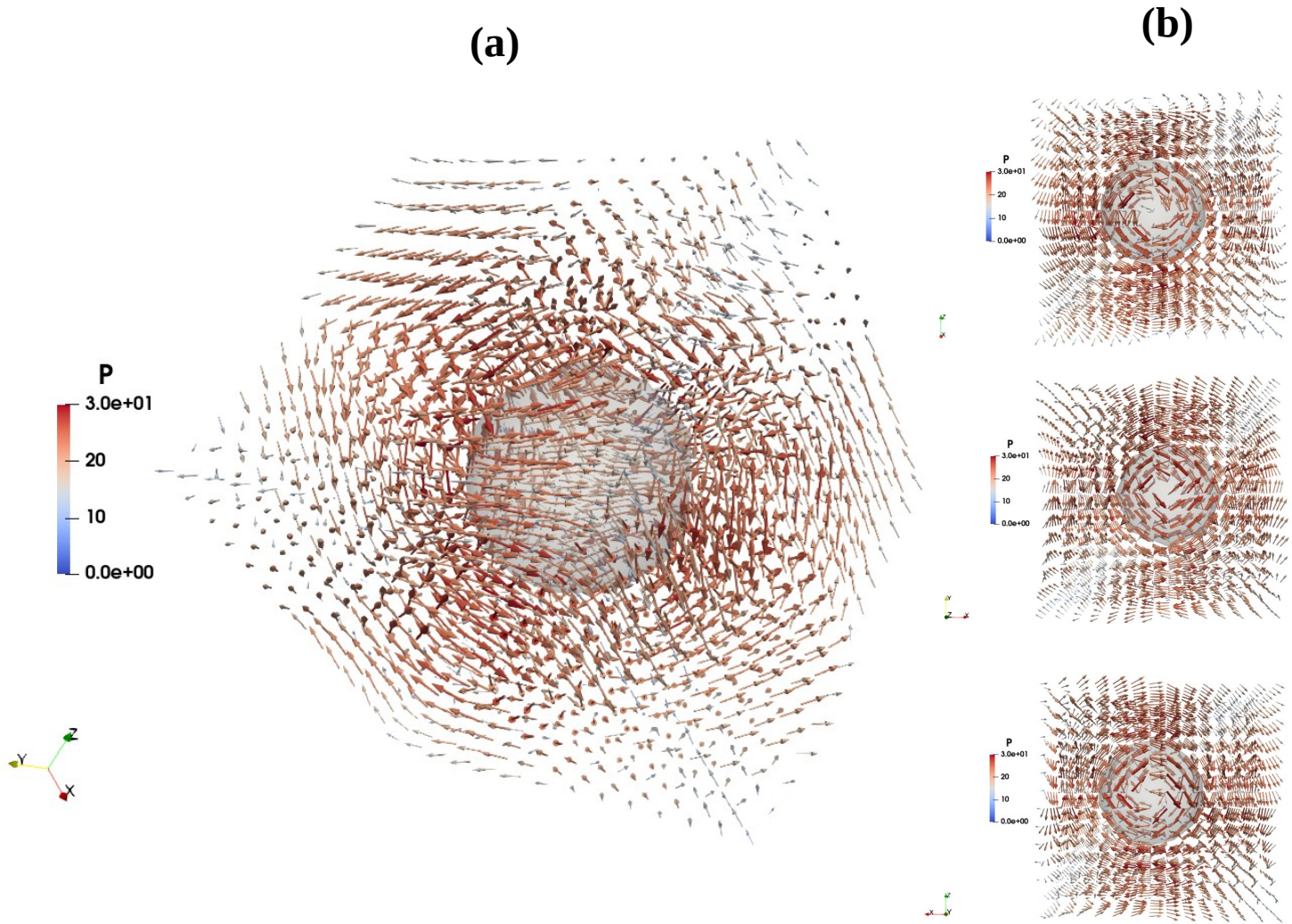

**FIG. S4**: Polarization pattern of the V$_6$ state stabilized in a BaTiO$_3$ matrix embedded with spherical BaZrO$_3$ nanoregions arranged in a (7,7) array, where the first digit denotes the diameter of the inclusions and the second their separation. (a) Polarization patterns around a single spherical inclusion. (b) Different views highlighting the independent swirling vortices converging at the dielectric nanosphere.

## 5. Raw data for the temperature dependence of P and G for the (5,5) array

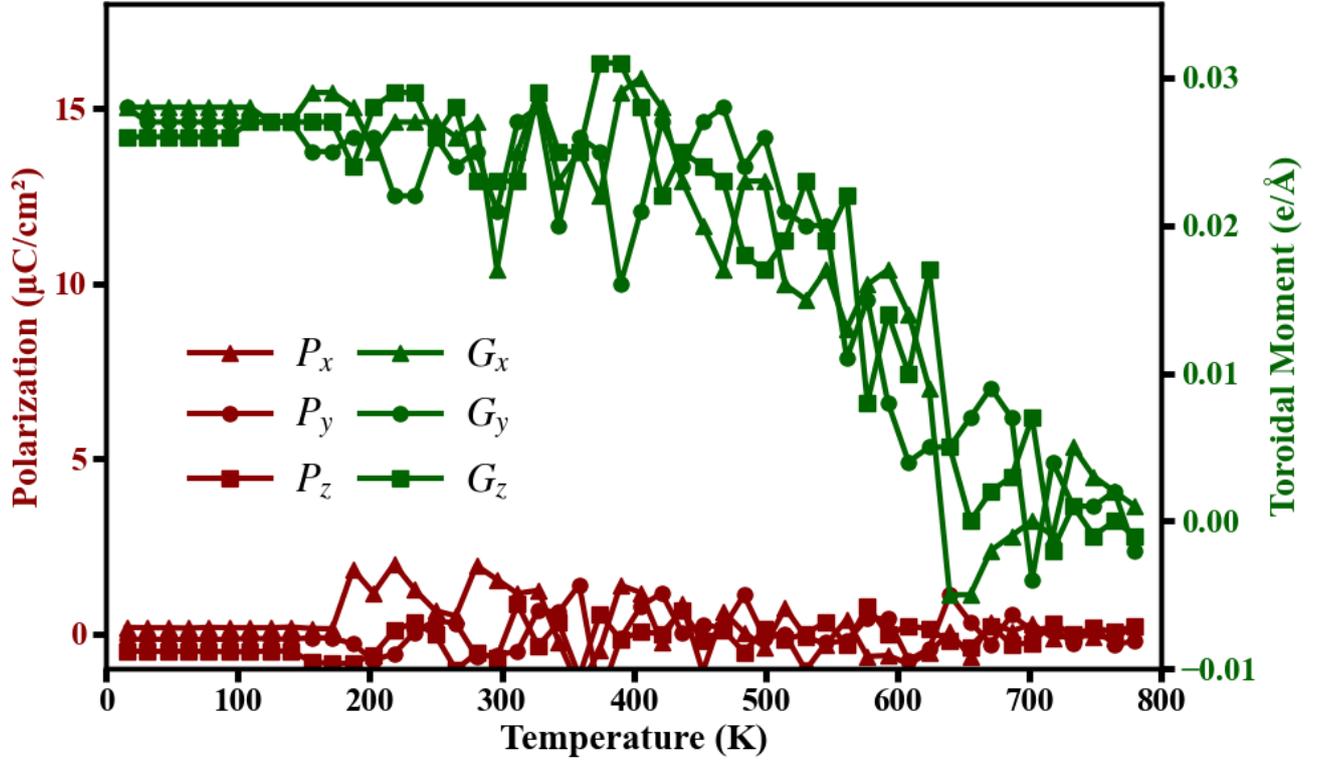

**FIG. S5**: Raw data of the temperature dependence of polarization P and toroidal moment G for the (5,5) array, illustrating the stabilization of the $V_6$ vortex supercrystal state. Note that the energy-window averaging technique described in the main text was applied to these raw datasets to produce the smoothed results presented in Fig. 3a.